\documentclass{PoS}
\usepackage{amsmath}
\usepackage{float}
\usepackage{comment}
\usepackage{amssymb}
\usepackage{color}
\usepackage[left]{lineno}
\usepackage{nameref}
\usepackage{amstext}
\usepackage{textcomp}

\newcommand*{\fig}{{\sc Fig.}}

\newcommand*{\tab}{{\sc Tab.}}
\usepackage{float}

\title{H.E.S.S. observations of very-high-energy emission from 1RXS J023832.6-311658}

\ShortTitle{1RXS J023832.6-311658}

\author{\speaker{Florian Gat\'e}\\
        Laboratoire d'Annecy-le-Vieux de physique des particules, 9 Chemin de Bellevue, 74940 Annecy-le-Vieux, France \\
        E-mail: \email{florian.gate@lapp.in2p3.fr}}

\author{for the H.E.S.S. collaboration}

\author{Thomas Fitoussi\\
       9, avenue du Colonel Roche, BP 44346, 31028 Toulouse Cedex 4, France \\
        E-mail: \email{thomas.fitoussi@irap.omp.eul}}

\abstract{Observations of the HBL blazar 1RXS J023832.6-311658 were made in 2013, 2015 and 2016 with the High Energy Stereoscopic System (H.E.S.S.). An excess of very high energy (VHE: > 100 GeV) gamma rays is clearly observed. The spectral energy distribution including the VHE spectrum will be presented. This object has a hard spectrum at TeV energies, and has a redshift z=0.23. These characteristics could be suitable for extragalactic magnetic fields for which the perspectives with H.E.S.S. will be discussed.}

\FullConference{35th International Cosmic Ray Conference €" ICRC2017-\\
		10-20 July, 2017\\
		Bexco, Busan, Korea}

\begin{document}

\section{H.E.S.S. observations}

H.E.S.S. (High Energy Stereoscopic System), located in the Khomas Highland in Namibia, is an array of five imaging atmospheric Cherenkov telescopes (IACT). It observes the gamma-ray sky with energies from tens of GeV up to around 100 TeV.

1RXS J023832.6-311658 is a high-frequency peaked BL Lac (HBL) object at z = 0.2329. It has never been detected by Imaging Atmospheric Cerenkov Telescopes (IACTs) before its observation by the H.E.S.S. phase II array in September and November 2013. In this contribution, we present the discovery of VHE emission from 1RXS J023832.6-311658 and its characteristics.

For the analysis, 15.8 hours of high quality data, after acceptance correction, taken in 2013, 2015 and 2016 with H.E.S.S II were used. The data were analyzed with the Model++ analysis \cite{Model} using standard cuts and cross-checked with the ImPACT analysis \cite{ImPACT}. The results from both analysis methods are in good agreements. 

\begin{figure}[h]
 \begin{center}
 \includegraphics[scale=0.4]{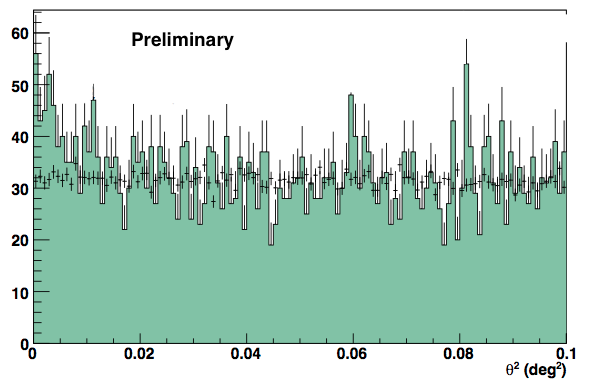}
  \includegraphics[scale=0.98]{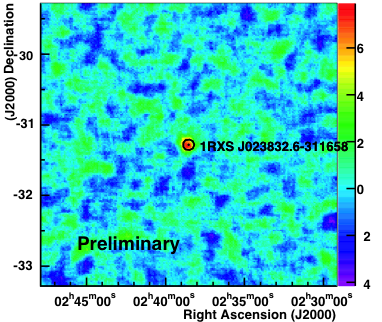}
   \caption{Left: $\theta^2$ distribution of the photons. Right: H.E.S.S II significance map obtained with the ring background method calculated from the Li and Ma formula \cite{LiMa}.}
\label{fig:detection}
  \end{center}
\end{figure}

The \fig~\ref{fig:detection} (left) shows the on-source and normalized off-regions distributions as a function of  squared angular distance ($\theta^2$) to target position. The map of the photon excess significance, shown on the right panel of \fig~\ref{fig:detection} was calculated with the Li and Ma formula \cite{LiMa}. 

1RXS J023832.6-311658 is clearly detected with a 5.9 standard deviation ($\sigma$) statistical significance with no other significant excess towards the target position.

\section{Source characteristics}

\begin{figure}[H]
 \begin{center}
 \includegraphics[scale=0.4]{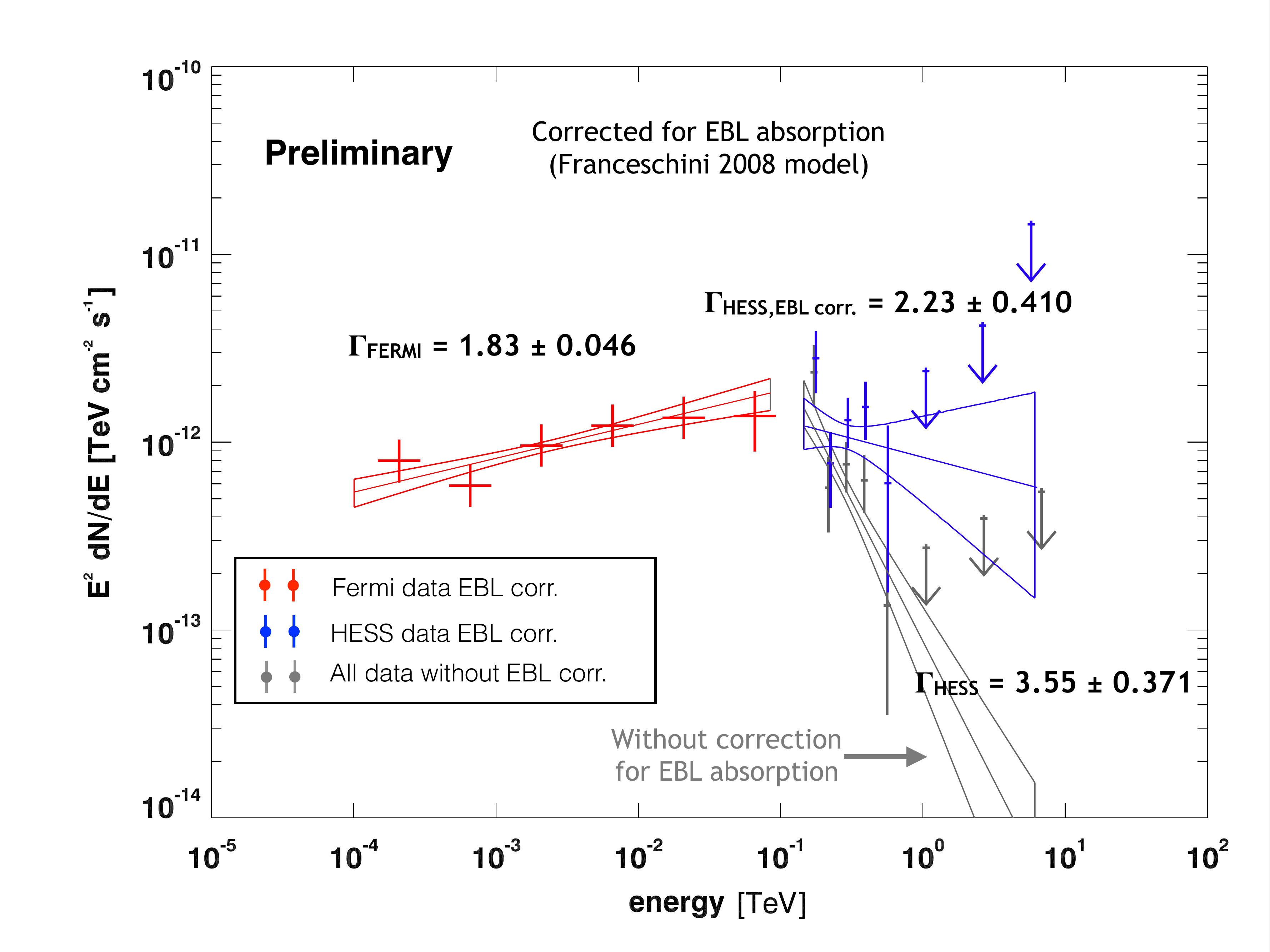}
   \caption{High energy 1RXS J023832.6-311658 spectrum. H.E.S.S and FERMI data are displayed together}
\label{fig:spectre}
  \end{center}
\end{figure}

The VHE photon spectrum is shown in \fig~\ref{fig:spectre}, in which HESS-II and FERMI-LAT spectra, corrected for EBL absorption using the Franceschini 2008 model \cite{frans}, are displayed. The behaviors of both data sets are well described by power-law functions $\text{d}N/\text{d}E = N_0 (E/E_0)^{-\alpha}$ whose parameters are summarized in \tab~\ref{tabPL}. One can notice that 1RXS J023832.6-311658 emits VHE photons with a hard spectrum, characterized by a power law index $\alpha = 2.232 \pm 0.410$

\begin{figure}[H]
 \begin{center}
 \includegraphics[scale=0.4]{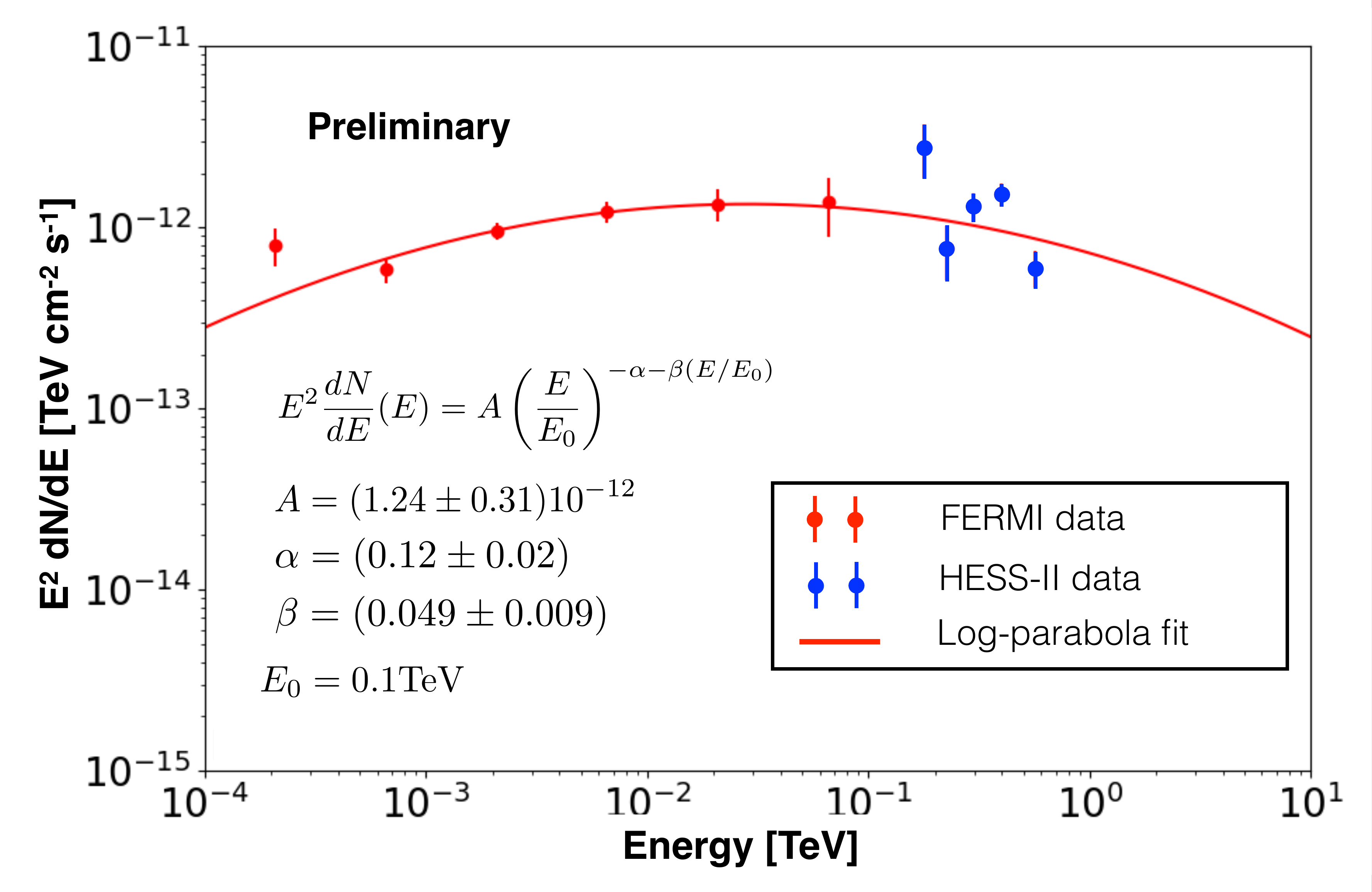}
   \caption{1RXS J023832.6-311658 Spectrum modeling according to H.E.S.S and FERMI data. The fit is performed using a log parabola model}
\label{fig:modspectre}
  \end{center}
\end{figure}

Several models have been fit to the full high energy data set (HESSII + FERMI-LAT). A goodness-of-a-fit $\chi^2$ test was performed to find the model that best describes the data. The best agreement was obtained for a Log-parabola model, as shown in \fig~\ref{fig:modspectre}, for which we obtain $\chi^2$/d.o.f. = 5.3 ($\chi^2$/d.o.f. = 38.9 for a power-law model). The HESS II data, combined with the FERMI-LAT data, allow to constrain the position of the gamma-ray peak. After fitting the data, the peak location is estimated to be at an energy of 32.3 $\pm$ 1.2 GeV, which is consistent with
general HBL spectra morphologies. The error on the peak position is derived from the errors on the fit free parameters, ($A$, $\alpha$ and $\beta$).

\begin{figure}[H]
 \begin{center}
 \includegraphics[scale=0.4]{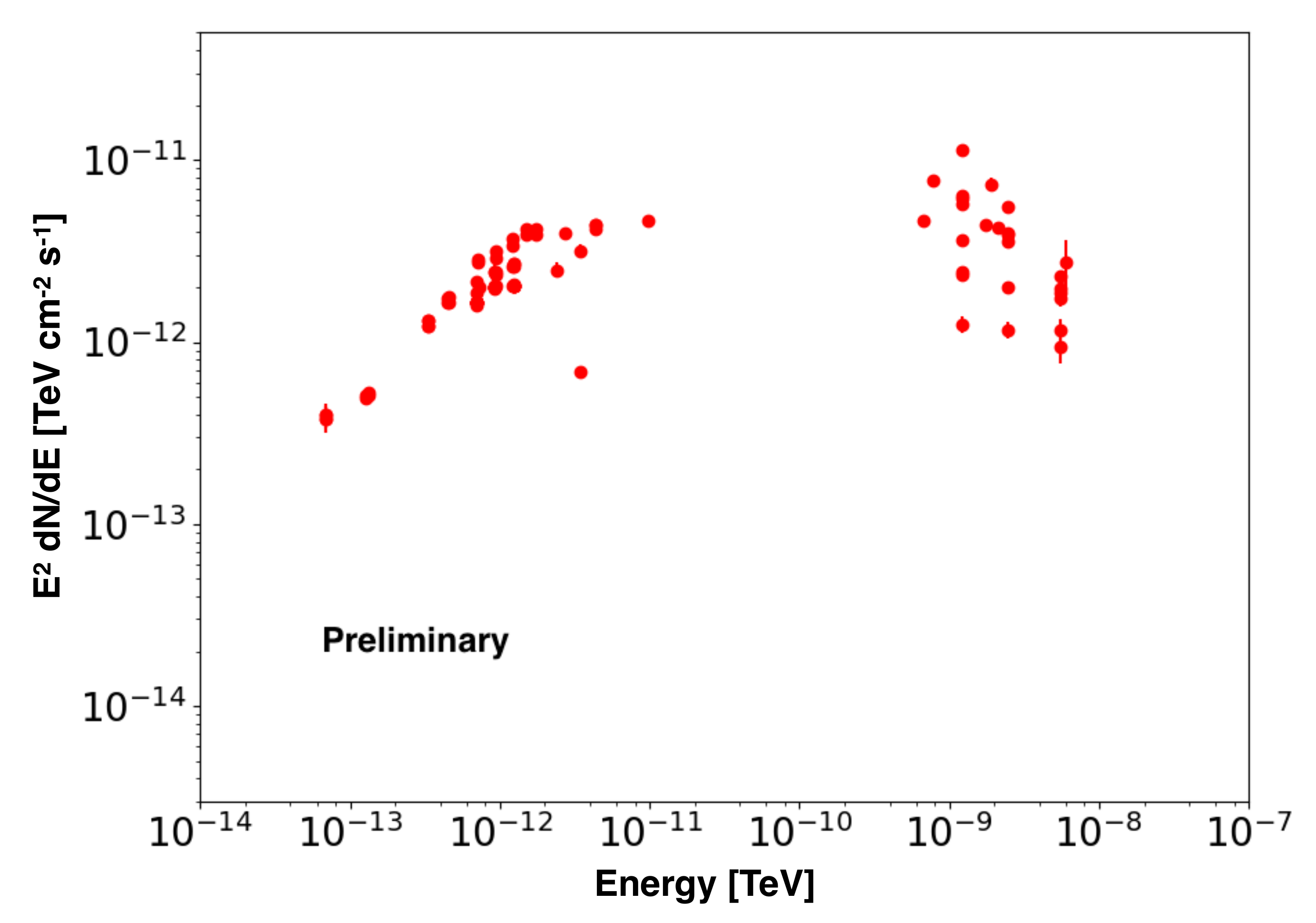}
   \caption{Low energy spectrum}
\label{fig:lowspectre}
  \end{center}
\end{figure}

1RXS J023832.6-311658 has also been observed by multiple experiment at lower energies, as shown in \fig~\ref{fig:lowspectre}, where only data after the year 2000, available from the ASI Science Data Center (ASDC) \cite{ASDC}, are displayed. The synchrotron peak position seems to be located around 100 eV. However, the lack of data due to the galactic neutral absorption in the energy range [10 eV - 1 keV] do not allow to constrain the synchrotron peak energy precisely.

\begin{table}
\begin{center}
\resizebox{\textwidth}{!}{
\begin{tabular}{|l|c|c|c|c|c|}
        \hline
   Instrument & energy range [GeV] & $N_0$ [TeV.cm$^{-2}$.s$^{-1}$] & $E_0$ [TeV] & $\alpha$ & $\beta$  \\ \hline \hline
   FERMI Lat  & 0.1 - 100 & (10.486 $\pm$ 0.118 )$\times 10^{-12}$& 0.68 & 1.834 $\pm$ 0.046 & -  \\
   HESS II without EBL correction &   > 100  & (7.448 $\pm$ 1.090) $\times 10^{-12}$ & 0.288 & 3.552 $\pm$ 0.371 & - \\
    HESS II with EBL correction &   > 100  & (12.945 $\pm$ 1.896) $\times 10^{-12}$ & 0.288 & 2.232 $\pm$ 0.410 & - \\
   FERMI Lat + HESS II  & > 0.1 & $(1.24 \pm 0.31) \times 10^{-12}$  & 0.1 & 0.12 $\pm$ 0.02 & 0.049 $\pm$ 0.009  \\ \hline
\end{tabular}}
\end{center}
\caption{Summary of the power-law function describing the HESS-II and FERMI-LAT spectra, the last line gives the parameters of the Log-parabola function modelling the full data set (HESS-II and FERMI-LAT data).}\label{tabPL}
\end{table}

\begin{figure}[H]
 \begin{center}
 \includegraphics[scale=0.4]{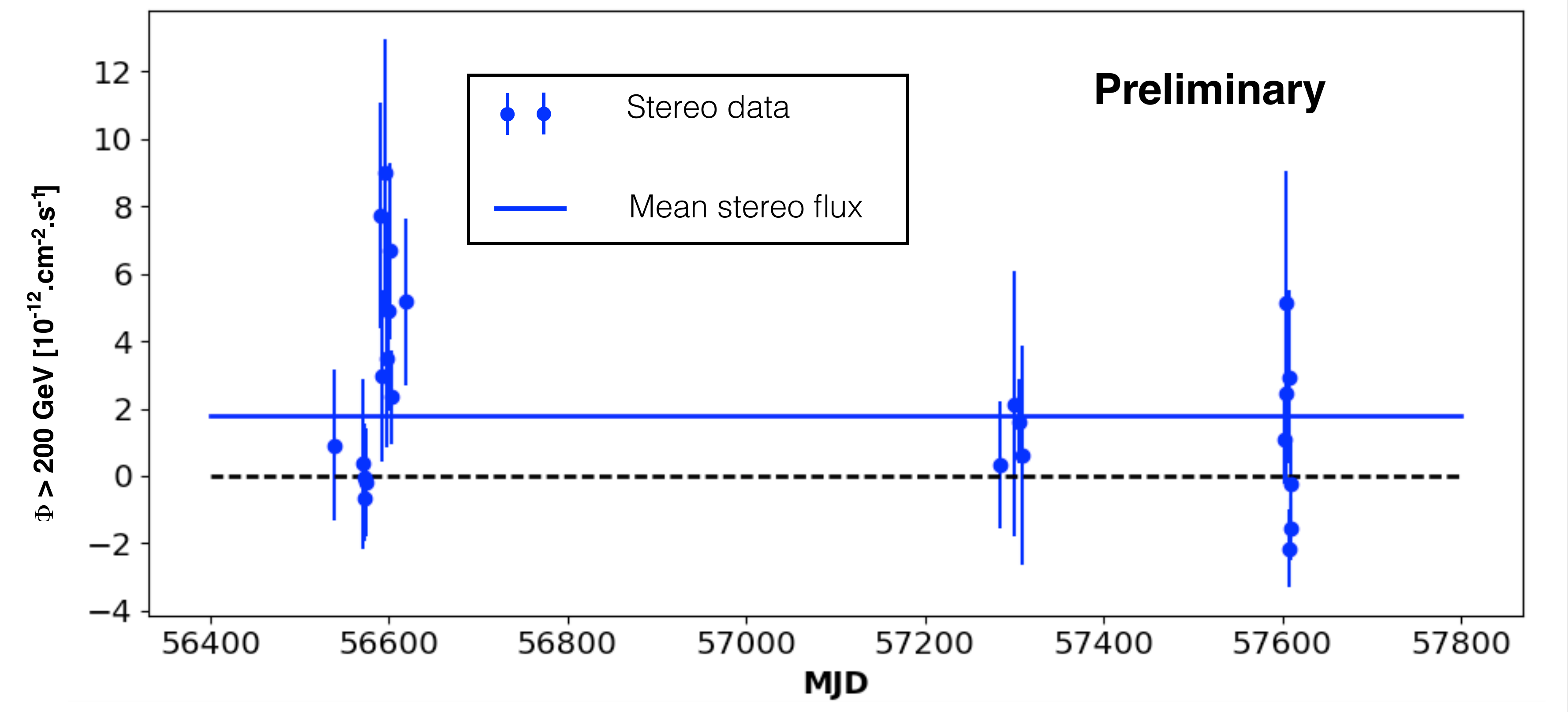}
   \caption{Night by night light curve.}
\label{fig:lightnight}
  \end{center}
\end{figure}

The night by night light curve is shown on \fig~\ref{fig:lightnight}. The mean flux over the three observation periods is $\phi = 1.79 \times 10^{-12}$ cm$^{-2}$.s$^{-1}$ over 200 GeV. To quantify the emission variability, a $\chi^2$ statistical test is performed to test the constant emission hypothesis. We obtain a $\chi^2 = 59.78$, the degrees of freedom (dof) = 23, leading to a p-value p = 0.0001. As a consequence the steady emission hypothesis is rejected.

In the next section, we present a preliminary study where we simulate IGMF effects on the gamma rays propagation from 1RXS J023832.6-31165. The object is highly likely to be variable, but for simulation purpose we assume that the source has a low variability emission.


\section{Possibilities of IGMF studies with 1RXS J023832.6-311658}

\begin{figure}[H]
 \begin{center}
 \includegraphics[scale=0.4]{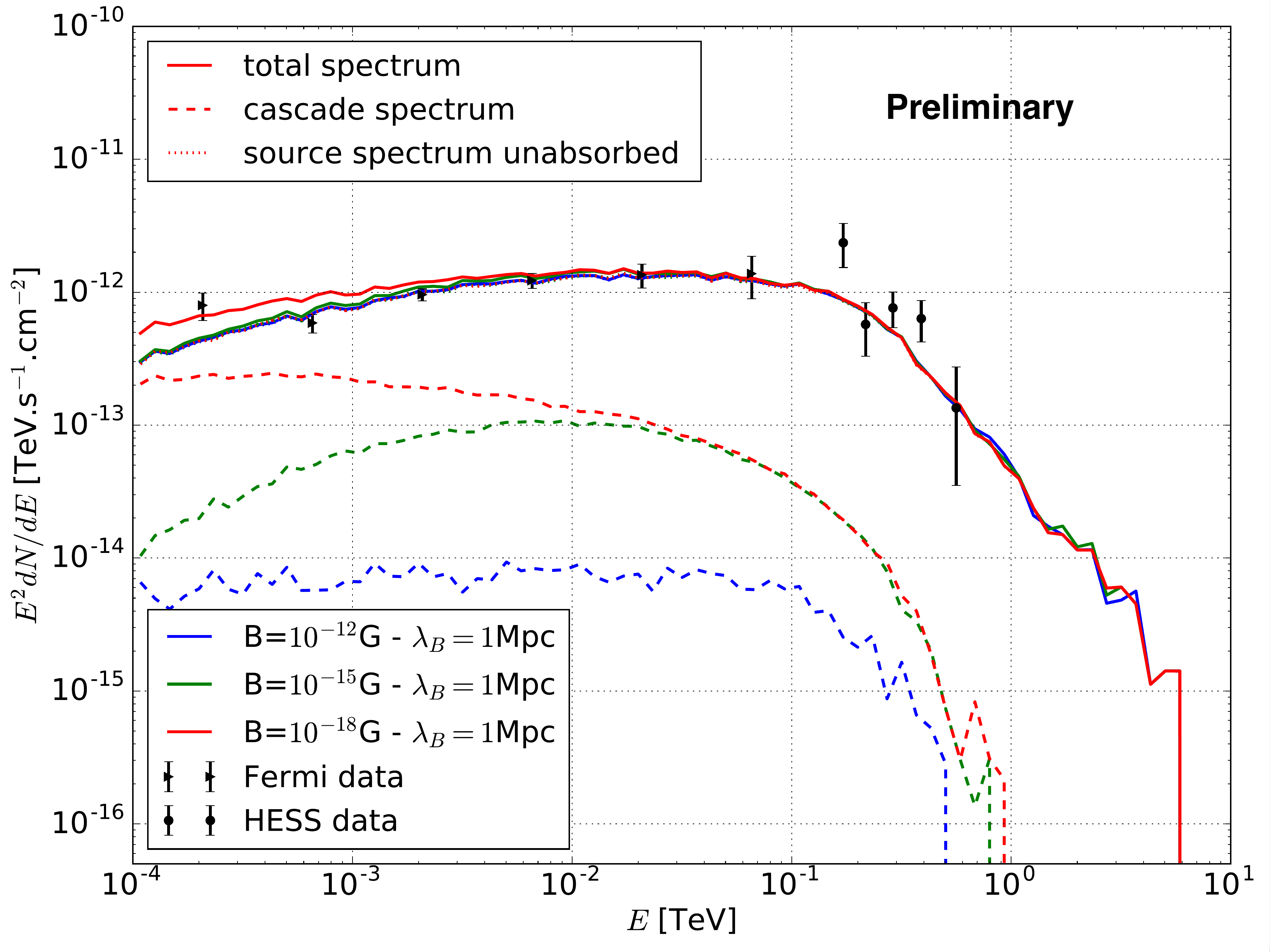}
   \caption{Electromagnetic cascade contribution to the observed spectrum, computed with \cite{FITOUSSI}.}
\label{fig:cascade}
  \end{center}
\end{figure}

Fitting the Fermi-LAT spectral points (so assuming they are intrinsic and not affected by IGMF effects), together with the H.E.S.S. points using a log-parabola between 100 MeV and 10 TeV, we compute the cosmological electromagnetic cascade contribution with the code presented in \cite{FITOUSSI}. Results are shown in \fig~\ref{fig:cascade}. Simulations have been done for three values of the intergalactic magnetic field (IGMF): $B=10^{-12}$ G, $B=10^{-15}$ G, $B=10^{-18}$ G. Coherence length $\lambda_B$ has been set to 1 Mpc. Selection of photons inside a field of view of $0.5^{\circ}$ around the source have been made. The extragalactic background light model used is \cite{frans}.

The contribution of the cascade to the spectrum in a limited field of view is dependent on the EGMF amplitude \cite{dermer2011}. Simulations show that IGMF with an amplitude higher than $B=10^{-15}$ G, cascade does not contribute significantly to the total spectrum. Then this results are compatible with no cascade because the spectrum is totally generated by the source. Lower value of the EGMF amplitude can not be interpreted so easily. For the moment the intrinsic spectrum used to compute the cascade is fitted with the data over the full range of energy (100 MeV - 10 TeV). Since the cascade contributes to the spectrum below 1 TeV (see dashed curve figure 6), data points used to fit this spectrum can already contain a contribution of the cascade and is not only the intrinsic spectrum. Thus using this spectrum to compute the cascade can lead to overestimate the cascade contribution. The intrinsic spectrum modeling in the simulations will be refined before putting limits on the IGMF. 

This preliminary result shows that this object is of moderate interest to probe the IGMF. Nevertheless, detecting more sources with moderate redshift is of importance to better understand the intergalactic medium.

\end{document}